\documentclass[sigconf,authorversion]{acmart}

\usepackage{tikz}
\usetikzlibrary{positioning}

\begin{document}
\title{Learning from Uncurated Regular Expressions \\ for Semantic Type Classification}

\author{Michael Mior}
\orcid{0000-0002-4057-8726}
\affiliation{%
  \institution{Data Unity Lab, Rochester Institute of Technology}
  \streetaddress{102 Lomb Memorial Drive}
  \city{Rochester}
  \state{New York}
  \country{USA}
  \postcode{14623-5608}
}
\email{mmior@mail.rit.edu}

\copyrightyear{2023}
\acmYear{2023}
\setcopyright{acmlicensed}\acmConference[SiMoD '23]{1st Workshop on
Simplicity in Management of Data}{June 23, 2023}{Bellevue, WA, USA}
\acmBooktitle{1st Workshop on Simplicity in Management of Data (SiMoD '23),
June 23, 2023, Bellevue, WA, USA}
\acmPrice{15.00}
\acmDOI{10.1145/3596225.3596226}
\acmISBN{979-8-4007-0783-4/23/06}
\begin{abstract}

Significant work has been done on learning regular expressions from a set of data values.
Depending on the domain, this approach can be very successful.
However, significant time is required to learn these expressions and the resulting expressions can become either very complex or inaccurate in the presence of dirty data.
The alternative of manually writing regular expressions becomes unattractive when faced with a large number of values that must be matched.

As an alternative, we propose learning from a large corpus of manually authored, but uncurated regular expressions mined from a public repository.
The advantage of this approach is that we are able to extract salient features from a set of strings with limited overhead to feature engineering.
Since the set of regular expressions covers a wide range of application domains, we expect them to be widely applicable.

To demonstrate the potential effectiveness of our approach, we train a model using the extracted corpus of regular expressions for the class of semantic type classification.
While our approach yields results that are overall inferior to the state-of-the-art, our feature extraction code is an order of magnitude smaller, and our model outperforms a popular existing approach on some classes.
We also demonstrate the possibility of using uncurated regular expressions for unsupervised learning.

\end{abstract}

\maketitle

\section{Introduction}

Regular expressions are useful for recognizing a set of strings corresponding to a regular language.
In practice, regular expressions are useful to check whether a string conforms to a simple set of syntactic rules.
For example, the rule that an email address must contain the symbol \texttt{@} can be represented by \texttt{\^{}(.+)@(.+)\$}.
Such an expression is easy to write for someone familiar with regular expressions.
Although there are more accurate expressions for detecting email addresses, they also become increasingly complex.
Furthermore, as the number of patterns to match grows, so does the effort to generate efficient and accurate regular expressions.
One approach to solving this problem is regular expression learning, which we discuss in more detail in Section~\ref{sec:related}.
A primary drawback to regular expression learning is the time it takes to learn the expressions.

Since regular expressions are challenging to write, we instead propose using an existing corpus of regular expressions and allow the learning process to use these expressions to extract useful features from text.
We provide more details on our proposed approach in the following section.

\section{Uncurated Expression Learning}

We propose an alternative approach to learning from regular expressions by making use of a large dataset of manually crafted regular expressions.
Specifically, we use regular expressions from regex101~\cite{Regex101}, a website that allows users to write, test, and, more importantly, save regular expressions.
These expressions may be for any purpose the user desires, but importantly, we assume that many of these expressions were crafted by users to learn \emph{something} meaningful.
We used the regex101 API to collect all expressions that were saved to the site as of March 2023 which resulted in a total of 16,707 expressions.

Many of these expressions appear to be useless.
For example, several hundred expressions have length 4 or less and it is unlikely that expressions of this length represent anything meaningful.
In contrast, several expressions in the dataset are tens of thousands of characters long, which we also expect are unlikely to represent anything of value.
Since one goal of regex101 is to assist users in learning to write regular expressions, we expect several of the expressions to be incorrect or simply invalid.
Some sample expressions we have collected are shown in Figure~\ref{fig:examples}.

\begin{figure}[ht]
\raggedright
$RE_1:$ \texttt{(\textbackslash{}d*)( *)(x|\textbackslash{}*)} \\
$RE_2:$\texttt{(public)*\textbackslash{}s*class\textbackslash{}s+(\textbackslash{}S+)\textbackslash{}s+} \\
$RE_3:$\texttt{\^{}[0-9]*(\.[0-9]*)*?(-[\^{}\textbackslash{}s]*)?\$} \\
$RE_4:$\texttt{\^{}([<>=!]?)\textbackslash{}s?((?:\textbackslash{}d*\textbackslash{}.)?\textbackslash{}d+)\textbackslash{}s?(.*)\$} \\
$RE_5:$\texttt{(https)(.*?)(720P)(.*?)(\textbackslash{}n)} \\
$RE_6:$\texttt{([a-z]|[A-z]){3,5}\#(H|F|Z|Q)([1-2]|[3-4]|[5-6]|[7-8])} \\
$RE_7:$\texttt{(\textbackslash{}w+):\textbackslash{}s*(\textbackslash{}w*)} \\
$RE_8:$\texttt{\textbackslash{}S+@gmail.com} \\
$RE_9:$\texttt{(d)(?:ays?)?|(w)(?:eeks?)?} \\
$RE_{10}:$\texttt{\^{}UTC[+-]\textbackslash{}d\{2\}[:]\textbackslash{}d\{2\}\$} \\
$RE_{11}:$\texttt{228-1234567} \\
$RE_{12}:$\texttt{\^{}(\textbackslash{}d\{4\}-\textbackslash{}d\{2\}-\textbackslash{}d\{2\})\$}
$RE_{13}:$\texttt{\^{}\textbackslash{}\{2\}(\textbackslash{}/|-)\textbackslash{}d\{2\}(\textbackslash{}/|-)(\textbackslash{}d\{2\}|\textbackslash{}d\{4\})\$}
    \caption{Example regular expressions}\label{fig:examples}
\end{figure}

Some of these expressions are unlikely to be useful, such as $RE_{11}$ which matches a specific literal string.
This is expected due to the uncurated nature of these expressions.
However, we also see other expressions that we expect to be useful.
$RE_{10}$ is capable of matching a certain class of strings representing time zones, while $RE_8$ will match a significant set of email addresses.
Although each of these expressions may not be of significant utility individually, we expect that by leveraging the entire corpus, we can still extract useful information.

For example, $RE_{12}$ and $RE_{13}$ both recognize dates in different formats.
When learning regular expressions, it can be difficult to learn expressions that capture these multiple formats without creating expressions that are unnecessarily complex.
In this case, by using a large corpus of expressions, we are able to extract features that can capture these different formats without additional effort.

We note that this of course requires a relevant expression to be contained in the corpus, but with the broad set of domains covered in the corpus, we expect a wide variety of use cases to be covered.
Furthermore, if the set of expressions is insufficient, we can augment this approach with existing regular expression learning techniques.
That is, we can apply regular expression learning to any values not correctly matched and add those to the corpus.
This allows a reduction of learning time without sacrificing the flexibility to learn expressions on arbitrary data.

\section{Preliminary Results}

We explore two possible applications for uncurated regular expression learning.
Firstly, we examine the possibility of using features extracted from our corpus of regular expressions as input to a machine learning model to solve the semantic type classification problem.
We also consider the possibility of using these features to perform clustering using an unsupervised approach.

\subsection{Semantic Type Classification}

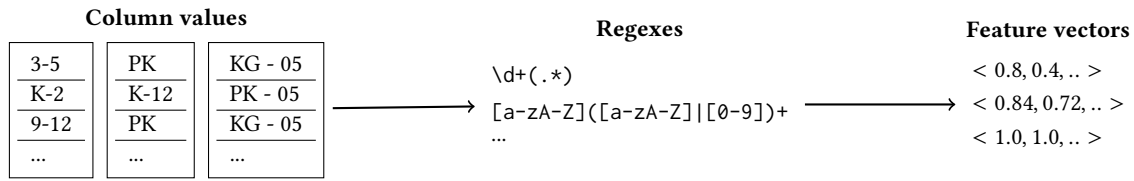
\begin{figure*}
    \centering
    \begin{tikzpicture}
    \matrix (cols) [ampersand replacement=\&, column sep=5] {
        \node (col1) [shape=rectangle,draw] {
            \begin{tabular}{l}
                3-5 \\\hline
                K-2 \\\hline
                9-12 \\\hline
                ...
            \end{tabular}
        };
        \& 
        \node (col2) [shape=rectangle,draw] {
            \begin{tabular}{l}
                PK \\\hline
                K-12 \\\hline
                PK \\\hline
                ...
            \end{tabular}
        };
        \&
        \node (col3) [shape=rectangle,draw] {
            \begin{tabular}{l}
                KG - 05 \\\hline
                PK - 05 \\\hline
                KG - 05 \\\hline
                ...
            \end{tabular}
        }; \\
    };
    
    \node [above=0cm of cols] {\textbf{Column values}};
    
    \matrix [above right = -1.55cm and 2cm of col3] (regexes) {
    \node {\texttt{\textbackslash{}d+(.*)}}; \\
    \node {\texttt{[a-zA-Z]([a-zA-Z]|[0-9])+}}; \\
    \node {...}; \\
    };
    
    \node [above=0cm of regexes] {\textbf{Regexes}};
    
    \draw [->, thick] (cols) edge (regexes);
    
    \matrix [right = 0 and 2cm of regexes] (vecs) {
    \node {$<0.8, 0.4, ..>$}; \\
    \node {$<0.84, 0.72, ..>$}; \\
    \node {$<1.0, 1.0, ..>$}; \\
    };
    
    \node [above=0cm of vecs] {\textbf{Feature vectors}};

    \draw [->, thick] (regexes) edge (vecs);
    
    \end{tikzpicture}
    \caption{Feature extraction using regular expressions}
    \label{fig:feature_extraction}
\end{figure*}

One task we expect our learning techniques to be useful is for semantic type classification.
The goal of semantic type classification is to add semantically meaningful annotations to set of values that correspond to a semantic type class, such as an entity type in a knowledge graph.
Examples of semantic types include place names, email addresses, and ages of persons.
Note that these types provide significantly more information than primitive types, such as strings or numbers.

We make use of data from the VizNet corpus~\cite{Hu2019} as annotated with 78 semantic types by the authors of Sherlock~\cite{Hulsebos2019}.
The dataset consists of sets of column values which are each annotated with a semantic type.
As stated above, our goal is to train a model that, given a set of column values, can predict the semantic type of that column.

To test the feasibility of uncurated regular expression learning for this approach, we set up a simple feature extraction pipeline as shown in Figure~\ref{fig:feature_extraction}.
For each set of column values, we match each value in the column against a large number of regular expressions from our corpus.
This matching step makes use of Hyperscan~\cite{Wang2019} to accelerate the simultaneous matching of regular expressions.
The result is a feature vector for each column where each element of the vector represents the fraction of values in the column that matched the expression.

For our preliminary model, we included all expressions in the Perl Compatible Regular Expressions (PCRE) dialect in the set collected from regex101 that are supported by Hyperscan, consisting of 6,538 expressions.
We note that after PCRE, the next most common regular expression dialect used by regex101 is that of the Javascript language.
Exploring the use of additional regular expressions is left for future work.
However, since our preliminary analysis covers almost half of the collected regular expressions, we do not expect a significant change in results.

An interesting observation as seen in the example in Figure~\ref{fig:feature_extraction} is that the collected regular expressions are able to extract salient features from the column values.
For example, the first regular expression will serve to extract the percentage of column values that contain digits, while the second matches strings consisting of letter followed by one or more letters or digits.
Individually, each of these features may perform poorly in correctly classifying our semantic types.
However, when taken as a whole, they can be effective training data.

\begin{table}[ht]
    \centering
    \begin{tabular}{l|l|l}
         Class & Ours & Sherlock \\\hline\hline
         address & 0.90 & 0.94 \\\hline
         birth date & 0.95 & 0.98 \\\hline
         director & 0.27 & 0.57 \\\hline
         gender & \textbf{0.82} & 0.79 \\\hline
         grades & 0.98 & 0.99 \\\hline
         isbn & 0.99 & 0.99 \\\hline
         manufacturer & 0.51 & 0.84 \\\hline
         nationality & 0.57 & 0.78 \\\hline
         person & 0.34 & 0.66 \\\hline
         publisher & 0.40 & 0.89 \\\hline
         year & \textbf{0.96} & 0.95 \\\hline
    \end{tabular}
    \caption{F-1 scores for semantic type classification}
    \label{tab:compare}
\end{table}

After performing the feature extraction, we trained a 5-layer dense neural network on the extracted set of feature vectors.
This is similar to one of the four feature specific subnetworks used by Sherlock.
Overall, we achieve a support-weighted F-1 score of 0.75.
This is significantly lower than Sherlock, which achieves a score of 0.90.
However, we note that our model achieved a better score in predicting some semantic types, such as ``year'' and ``gender''.
We show a sample of classes and F-1 score compared to Sherlock in Table~\ref{tab:compare}.
Note that the classes where our model performs well are those with some regular structure, while our model performs poorly on irregular types with large domains such as ``person''.
Although Sherlock uses a set of hand-crafted features, our model is much simpler.
Our code for feature extraction is approximately 50 lines while Sherlock has over 500 lines of code for feature extraction and makes use of several additional data sources including pretrained word embeddings and paragraph vectors.
This provides hope that while the performance of our model is currently much lower, there are promising uses for the set of features we extract.

This process of feature extraction can also be used to \emph{supplement} the set of features used by other models.
Currently, the features used by Sherlock include information on character distributions, word embeddings, paragraph vectors, and some global statistics.
None of these features captures patterns within column values.
We expect such features to be particularly useful when dealing with semantic types that have regular patterns such as email addresses and URLs.
We leave the integration of these features into systems such as Sherlock as future work.

We also note that the evaluation of Sherlock compared their approach with one using regular expression learning~\cite{Bartoli2016}.
The learned regular expression based on the provided examples produced an overall F1 score of only 0.04, suggesting that a single regular expression is not sufficient for this classification task.
Our approach of using the matches from multiple regular expressions however demonstrates that we can still leverage the ability of these expressions to detect patterns in the data useful for classification.

\subsection{Clustering}

Although our formulation of the semantic type classification above requires a ground truth, we also consider the possibility of grouping values into semantic types without requiring supervision.
We start with the same set of extracted features that we used above and consider their embedding in a lower-dimensional space.

Specifically, we consider the use of UMAP~\cite{McInnes2018} for dimension reduction.
UMAP is useful for visualizing higher-dimensional data.
In our case, we use UMAP to produce a two-dimensional plot to visualize the regular expression feature vectors.

\begin{figure}
    \centering
    \includegraphics[scale=0.55]{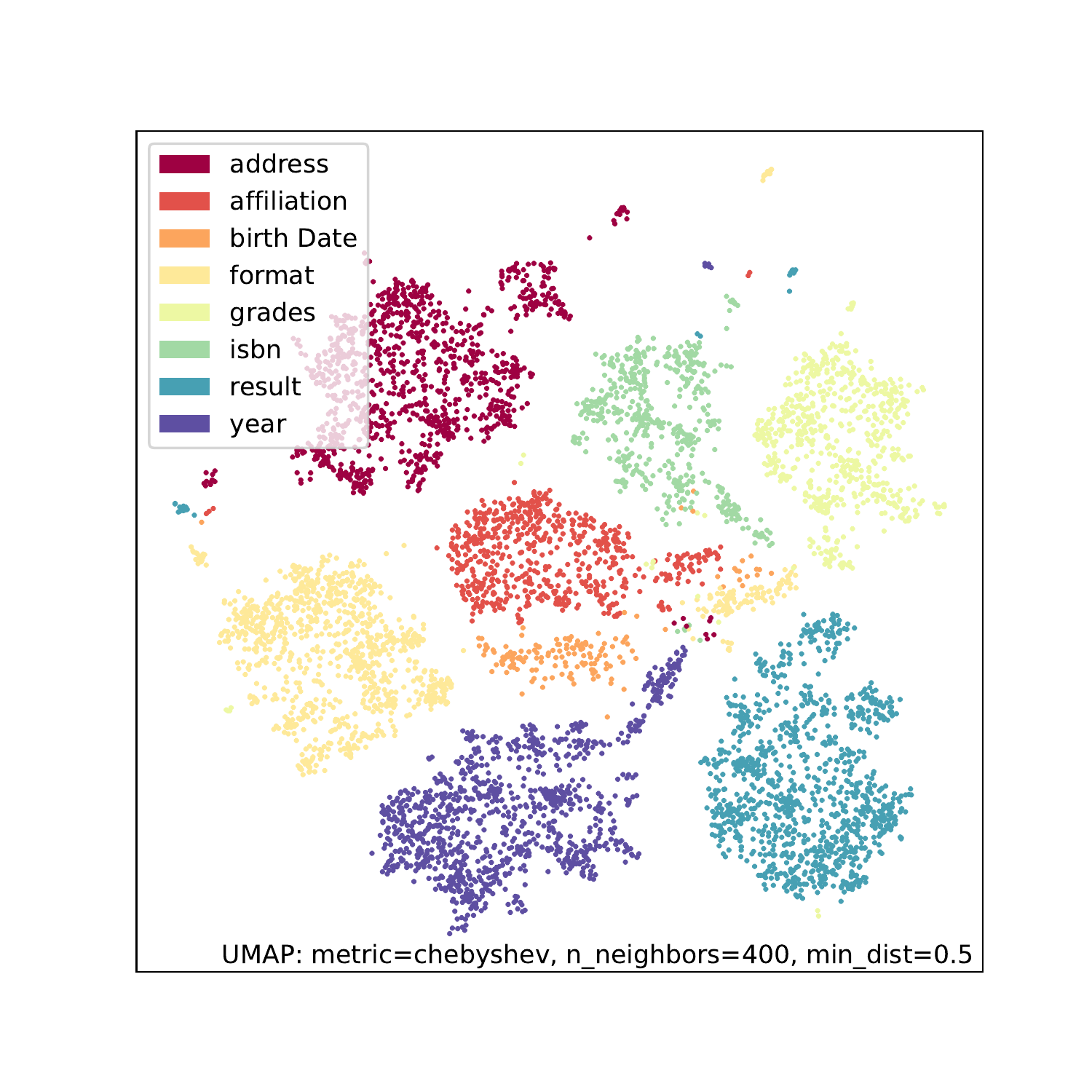}
    \caption{Two-dimensional UMAP transformation of feature vectors extracted using regular expressions}
    \label{fig:umap}
\end{figure}

We start with an embedding generated on our test data using all semantic classes with an F1 score greater than 0.9.
The embedding generated is shown in Figure~\ref{fig:umap}.
We can see that significant clusters appear in the dataset corresponding to groups of related values appearing together.
We believe that this suggests that dimensionality reduction techniques are promising in this setting.

Clustering of the extracted regular expression feature vectors is a suitable avenue for exploring unsupervised learning.
This has the advantage of not requiring ground truth labels.
Since our corpus of regular expressions covers a variety of domains, this has the potential to enable semantic classification for a broader class of applications.
This would be particularly useful in data lake scenario where there may be a large number of unknown semantic classes.
To evaluate the potential for clustering, we applied the DBSCAN~\cite{Ester96} clustering algorithm to the transformed data in Figure~\ref{fig:umap}.
Taking into account the best possible match between the discovered clusters and the original labels, we end up with an accuracy of $\sim{}0.86$, suggesting that unsupervised clustering is indeed viable.

There are many other possible applications of these feature vectors in the context of clustering.
For example, entity resolution makes use of features extracted from the data to determine possible duplicates.
Specifically, we expect that our regular expression features could be useful as a blocking function~\cite{Li2020} that is used to identify groups of potentially similar values to reduce the computational overheard of duplicate detection.
With appropriate feature selection to reduce the number of expressions, these regular expressions could be useful for blocking by grouping together values that match similar sets of regular expressions.

\section{Related Work}\label{sec:related}

We explore two avenues of related work.
The first is the learning of regular expressions from data, which is somewhat the opposite of our proposal.
The second is the problem of semantic type classification that we have highlighted above, as we feel this provides an interesting use case for uncurated regular expression learning.

\subsection{Regular Expression Learning}

There has been significant existing work in learning regular expressions from a set of strings.
Several authors have explored the use of regular expression learning for information extraction.
Li et al.~\cite{Li2018} define the problem of learning regular expressions as starting with a user-specified expression and learning an improved expression that is better at information extraction on a set of labeled examples.
This, of course, requires a regular expression to start with, an input that may not always be available.
Requiring such an expression also limits the scalability of this approach to a large range of classes, since user input is required for each class.
It also requires that all classes be defined in advance, while our approach could be applied in an unsupervised setting.

Other approaches~\cite{Bartoli2016,Bex2010,Brauer2011} do not require an initial regular expression.
However, these existing approaches are only able to detect occurrences that they have seen before.
Although there have been attempts to avoid overfitting, the expressions found will necessarily be tied in some fashion to the specific set of examples used for training.
By contrast, since the expressions in the corpus we explore are created independently of the data being matches, we expect generalization to be less of a concern.

\subsection{Semantic Type Classification}

Semantic type classification aims to assign a semantically meaningful type to one or more values.
Database schemas traditionally assign types such as strings or integers to values.
Semantic types provide more detail with types such as ``birth place`` or ``country``.
Existing approaches to semantic type classification make significant use of feature engineering.
Sherlock~\cite{Hulsebos2019} defines a set of features and uses deep learning to perform semantic type classification on table columns.
Sato~\cite{Zhang2020} improves Sherlock by taking into account the context of surrounding columns.
While these approaches perform well, the feature extraction process has significant overhead, and the features used by both of these approaches are unable to make use of patterns within data values.

As we showed earlier, our approach generally performs well with data values that follow a regular pattern and has the potential to outperform existing approaches in these cases.
However, we expect our approach to be complementary to existing approaches since we know it detects regular patterns well.
That is, our approach performs well with classes whose values exhibit strong syntactic similarity, such as ``isbn''.
Other features such as word and paragraph embeddings used as features in Sherlock that work with natural language can complement these syntactic features for classes whose values exhibit strong semantic similarity such as ``gender''.

\section{Future Work}

While we have demonstrated an example where the use of a library of uncurated regular expressions can have some value using supervised learning, there are further opportunities for exploring unsupervised learning.
For example, the match vectors produced by the observed regular expressions could be useful in clustering values to identify possible types.

There is also a significant potential benefit to performing data cleaning before attempting to use the expressions for learning.
For example, a brief inspection shows that our set of expressions contains both \texttt{.} and \texttt{.*}.
Both of these expressions will match any non-empty string, making these expressions useless for classification.
There are also several expressions in the set that we expect to be effectively noise in that we would expect them to match only a few or no strings in practice.
(As mentioned earlier, several of the expressions we collect are tens of thousands of characters long.)

Many regular expressions in the dataset we collected were not used, as they originate from different regular expression engines.
It is possible that including a wider body of regular expressions could produce better results.
To that end, there are other possible sources of regular expressions that may be useful for our purposes.
Public source code repositories such as GitHub and Q\&A sites such as Stack Overflow contain many instances of code used for regular expression matching.
A similar technique was employed by Yan et al.~\cite{Yan2018} who used GitHub to extract functions that implement type detection logic.
This has the potential to broaden the domain of the set of expressions in the corpus, making it more generally applicable.

\section{Conclusions}

We presented a method for collecting a large corpus of regular expressions that we expect to be useful for various machine learning tasks.
Our preliminary classification model outperforms the state of the art on some classes while using far fewer features that are not specifically engineered.
We also demonstrated the potential usefulness of regular expression features for unsupervised clustering of data values into meaningful semantic type classes.
There are further promising applications using unsupervised learning that we have yet to fully explore.

\balance

\bibliographystyle{ACM-Reference-Format}
\bibliography{references}

\end{document}